\documentclass[aps,prb,twocolumn,showpacs]{revtex4}
\usepackage{amsmath}
\usepackage{bm}
\usepackage{graphicx}

\begin{document}

\title{Effective level attraction and magnetic flux-induced negative differential conductance in
two double quantum dot molecules embedded in an Aharonov-Bohm ring}

\author{M. L. Ladr\'on de Guevara$^1$, Gustavo A. Lara$^2$, and P. A. Orellana$^1$}
\address{$^1$Departamento de F\'{\i}sica, Universidad Cat\'{o}lica
del Norte, Casilla 1280, Antofagasta, Chile}
\address{$^2$Departamento de F\'{\i}sica, Universidad de Antofagasta, Casilla
170, Antofagasta, Chile}

\begin{abstract}
We study transport of non-interacting electrons through two quantum dot
molecules embedded in an Aharonov-Bohm interferometer.
The system in equilibrium exhibits bound states in the continuum (BIC) and total suppression of
transmission. It also shows a magnetic flux-dependent effective level attraction
and lines of perfect transmission when the intramolecular coupling
is weak. Out of equilibrium, the current displays two kind of negative differential
conductance (NDC) regions, which have different origins.
One is generated by the usual mechanism of the NDC arising in a double quantum dot system.
The other is induced by the magnetic flux, and it occurs at small voltages and for
a well definite range of the intramolecular couplings. We explain this effect in terms
of the level attraction displayed by the system.



\end{abstract}

\date{\today}
\maketitle

\section{Introduction}


Advances in experimental techniques
at the nanometer scale have allowed to realize and manipulate quantum dots in
a controlled way\cite{petta}. This has permitted to study in these systems
a wide spectrum of phenomena\cite{variousQD},
as well as to take advantage of some properties of quantum dots to build nano-devices,
such as rectifiers, amplifiers, lasers, and others \cite{ono,vidan,amplif,laser}.
For their similitude with atoms, quantum dots are often viewed as artificial atoms,
and two or more coupled quantum dots as artificial molecules.
Multiple quantum dot systems, and in particular the double quantum dot (DQD) molecule\cite{vanderwiel},
are of particular importance, because they are more controllable
than single quantum dots. Moreover, they admit different kind of connections to leads.
Initially, most of studies considered serial DQDs\cite{vandervaart}, but
posteriorly parallel\cite{holleitner,ghost,kang} and T-shaped \cite{kim, boese}
configurations were also examined,
emerging quantum interference effects due to the existence of different electronic paths.
Much attention has received the DQD molecule embedded in an Aharonov-Bohm ring, where
the presence of a magnetic flux adds a new tool in the control of the transport properties.
Aharonov-Bohm (AB) oscillations and Fano effect in a DQD molecule
embedded in a ring have been observed
experimentally\cite{holleitner,sigrist,ihn}.
There is much theoretical work supporting these
experiments and exploring new phenomena in the DQD embedded in a
ring. Fano resonances and the magnetic-flux
controllability of transport are examined, for example, in Refs.
\onlinecite{ghost,kang,fanodicke,moldoveanu,sztenkiel,dong}, both in
presence and the absence of electronic correlations. The existence
of bound states in the continuum (BICs) is
discussed in Refs. \onlinecite{ghost,fanodicke}. The interplay
between quantum interference and Kondo physics in the
parallel-coupled DQD has been also
explored\cite{ramsak,sztenkiel2}.
A closely related system which has received attention
is the AB interferometer with two embedded quantum dots\cite{kubala,apel}.
Kubala and K\"onig studied equilibrium transport in this
system, finding an \emph{effective flux-dependent level
attraction} in the linear conductance\cite{kubala}. This effect is
caused by the renormalization of levels by the leads.


From the applicability point of view, an interesting feature exhibited by the
transport through quantum dots is the negative differential conductance
(NDC)\cite{weis}.
NDC has been studied in single as well as in double quantum dot systems, and
it has applications in amplifiers and oscillators in the microwave,
mm-wave and Terahertz frequency ranges\cite{sollner}.
In multilevels quantum dots NDC can occur when states
have different couplings to the leads\cite{weinman,ciorga,thielmann,rogge}.
In a serial DQD, NDC can be produced
when the bias breaks the transmission channel extended along the
system\cite{aguado}.
Other theoretical works on generation of NDC in DQD connected in series are
Refs. \onlinecite{lara,fransson,wunsch,nguyen,pedersen}.
In a DQD embedded in an Aharonov-ring,
magnetic-flux-induced NDC was found in the strong interdot repulsion regime\cite{dong}.
A similar result was found by Mourokh and Smirnov in a DQD molecule with three terminals\cite{mourokh}.
Recently, it was reported NDC induced by the electronic correlation in a side-coupled
DQD\cite{lara2}.

In this work we study equilibrium and non-equilibrium transport
through two quantum dot molecules embedded in an Aharonov-Bohm
interferometer. We obtain analytical expressions for the
transmission and we calculate numerically the current at zero
temperature. In equilibrium, the transmission exhibits Fano
resonances, total reflection, and suppressed peaks as a
manifestation of bound states in the continuum. Moreover, we find
for small intramolecular couplings a flux-dependent effective
level attraction and lines of perfect transmission. This effect
occurs no matter how weakly coupled are the quantum dots
forming each molecule.

In the non-equilibrium regime, we identify two kind of NDC regions
in the $I$-$V$ characteristics, occurring at different scales and
of different origin.
One is generated by the usual mechanism of
the NDC in a double quantum dot systems, and it is independent of
the magnetic flux. The current will increase
or decrease with
voltage, depending on whether the voltage makes the levels of the
different quantum dots become aligned or not aligned. The
second NDC region is induced by the magnetic flux, and it occurs
only for a definite range of intramolecular couplings. An abrupt
rise of current occurs for small bias voltages as consequence of
the effective level attraction of the hybridized levels produced
by the magnetic flux. The decrease of current is result of the
destruction of this effect when the bias voltage is increased.

\section{Model}

\begin{figure}[t]
\begin{center}
  \includegraphics[width=9.8cm]{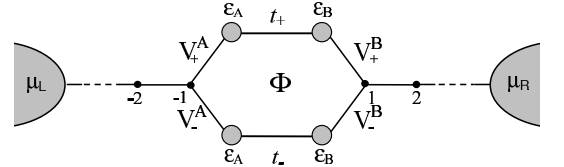}
  \end{center}
 \caption{Two double quantum dot molecules embedded in parallel in an Aharonov-Bohm
  interferometer}\label{Fig1}
\end{figure}

The system under consideration is shown in Fig. \ref{Fig1}.
Two equal double quantum dot molecules are embedded in an
Aharonov-Bohm
ring, which is attached to large contacts through one-dimensional leads.
Equilibrium transport in
a similar configuration with additional connections between dots was studied by
Li \emph{et al.}\cite{li}.
The left and right contacts are in thermodynamic equilibrium with
thermodynamical potentials $\mu_L$ and $\mu_R$, respectively.
The leads are assumed to be ballistic conductors.
We assume that a bias voltage $V/e$ is applied between source and drain such that
the site energy is $V/2$ for the left lead and $-V/2$ for the right lead.
We consider only one level relevant in each of the quantum dots.
The system is modeled by a non-interacting Anderson Hamiltonian, which can be
written as
\begin{equation} H=H_{M}+H_{0}+H_{I},  \label{eq-1}
\end{equation}
where $H_{M}$ describes the dynamics of the isolate molecules,
\begin{eqnarray}
H_{M}&=&\sum_{\alpha=+,-}\sum_{i=A,B}\varepsilon_{i}d_{i\alpha}^{\dag
}d_{i\alpha}\nonumber
\\ & &+ \sum_{\alpha=+,-}(t_{\alpha}
d_{A,\alpha}^{\dag }d_{B,\alpha}+t_{\alpha}^*d_{B\alpha}^{\dag
}d_{A\alpha}), \label{eq-2}
\end{eqnarray}
where $\varepsilon _{A(B)}$ is the level energy of the left
(right) quantum dot in the
molecule $\alpha$ ($\alpha=+,-$); $d_{i,\alpha}$ $%
(d_{i,\alpha}^{\dagger })$ annihilates (creates) an electron in
dot $i$ in the molecule $\alpha$, and $t_\alpha$ is the
intramolecular tunneling hopping.
$H_{0}$ is the Hamiltonian for
the noninteracting electrons in the leads,
\begin{equation}
\begin{split}
H_{0}=
&\;\frac{V}{2}\sum_{i=-N}^{-1}  c_{i}^{\dag}c_{i}+
v\sum_{i=-N}^{-1}(c_{i}^\dag c_{i-1}+c_{i-1}^\dag
c_i)\nonumber \\
& -\frac{V}{2}\sum_{i=1}^{N}
c_{i}^{\dag}c_{i} + v\sum_{i=1}^N(c_i^\dag
c_{i+1}+c_{i+1}^\dag c_i) \label{eq-3}
\end{split}
\end{equation}
where $c_{i}$ $(c_{i}^{\dag })$ is the annihilation (creation)
operator of an electron in the site $i$-th  of the leads, and
$v$ the hopping between sites in the leads. The term $H_{I}$
accounts for the tunneling between molecules and leads,
\begin{equation}
\begin{split}
H_{I} =
&-\sum_{\alpha=+,-} (V^A_\alpha d_{A\alpha}^{\dag}
c_{-1}+V^{A*}_\alpha c_{-1}^\dag d_{A\alpha}) \nonumber
\\
&- \sum_{\alpha=+,-} (V^B_\alpha d_{B\alpha}^{\dag}c_{1}+V^{B*}_\alpha c_{1}^\dag
d_{B\alpha})
\end{split}
\label{eq-4}
\end{equation}
with $V^{A(B)}_\alpha$, the tunneling coupling connecting the left
(right) dot of the $\alpha$-th molecule with the left (right)
lead. We restrict to the case in which there is not magnetic field acting directly
on the electrons, so that the situation will be identical for the two values of spin.
Therefore in what follows we omit the spin index.

In presence of a magnetic flux threading the ring, and using gauge invariance,
we add the Aharonov-Bohm phase $\phi=2\pi\Phi/\Phi_0$ around the ring by the replacement
\begin{eqnarray}
V^{A}_{+}&=& t_{A}e^{-i\phi/6}, \quad t_{+}=t_c e^{-i\phi/6} \quad
V^{B}_{+}=t_{B}e^{-i\phi/6},
\nonumber\\
V^{A}_{-}&=&t_{A}e^{i\phi/6}, \quad t_{-}=t_c e^{i\phi/6} \quad
V^{B}_{-}=t_{B}e^{i\phi/6}, \label{eq-5}
\end{eqnarray}
with  $\Phi_0=h/e$ the flux quantum.
We look for the steady states $|\psi_k\rangle$ of the whole
Hamiltonian $H$. The Hamiltonian describing the leads, $H_0$,
corresponds to a free-particle Hamiltonian on a lattice, the
eigenfunctions being Bloch functions
\begin{equation}
|k_\beta\rangle=\sum_j e^{ik_\beta j}|j\rangle,\quad \beta=L,R,
\end{equation}
where $|k_\beta\rangle$ is the momentum eigenstate and $|j\rangle$ a Wannier state localized at
the $j$-th site. The corresponding dispersion relations are
$\varepsilon=V/2-2v\cos k_L$, for the electrons originated in the left contact,
and $\varepsilon=-V/2-2v\cos k_R$ for those originated in the right contact.
The eigenstates of the entire Hamiltonian can be written as
\begin{equation}
|\psi_k\rangle=\sum_{j=-N}^{-1}a^k_j|j\rangle+ \sum_{\alpha=+,-}\sum_{i=A,B}b_{i,\alpha}^k|i,\alpha\rangle +\sum_{j=1}^N a^k_j|j\rangle
\label{eq-6}
\end{equation}
We assume electrons as described by a plane wave incident from the far left (right) with unit amplitude, reflection amplitude $r$ ($r'$), and transmission amplitude $t$ ($t'$).
Thus, for electrons incident from the left,
\begin{equation}
a_j^{k_L}=
\begin{cases}
e^{ik_Lj}+re^{-ik_Lj}   &  j\leq -1     \\
   te^{ik_Rj}, &    j\geq 1  \\
\end{cases}
\label{eq-7}
\end{equation}
and for electrons incident from the right,
\begin{equation}
a_j^{k_R}=
\begin{cases}
e^{-ik_Rj}+r'e^{ik_Rj}    &  j\geq 1     \\
 t'e^{-ik_Lj},  &    j\leq -1  \\
\end{cases}
\label{eq-8}
\end{equation}
where the two contributions are independent.
Inserting Eqs. (\ref{eq-6}),(\ref{eq-7}) and (\ref{eq-8}) in the
Schr\"odinger equation $H|\psi_k\rangle=E_k|\psi_k\rangle$, we can
solve for $a_j^k$ and $b_{i,\alpha}^{k}$.
We are interested in the transmission
and the current through the system for an applied voltage $V/e$ between contacts.
We center in the the symmetrical configuration, that is, equal left and right
dot-lead couplings, $t_A=t_B$. Additionally, the energies of the quantum dots are
$\varepsilon_A=\mu_L$ and $\varepsilon_B=\mu_R$.

The current in the leads is given by
\begin{equation}
I=\frac{2iev}{\hbar}\left(\langle c^\dag_{j+1}c_j\rangle
-\langle c^\dag_{j}c_{j+1}\rangle\right),
\label{eq-30}
\end{equation}
where
\begin{equation}
\langle c^\dag_{i}c_j\rangle =\frac{1}{2}\sum_{\alpha=\{L,R\}}\frac{1}{N}\sum_{k_\alpha}f(\varepsilon_{k_\alpha}-\mu_\alpha)a_i^{k_\alpha *}a_j^{k_\alpha}.
\label{eq-31}
\end{equation}
We assume that the voltage drop occurs only
between the dot $A$ and the dot $B$, so that the energies of the quantum dots are equal to
the site energies of the adjacent leads, $\varepsilon_{A}=V/2$ and
$\varepsilon_B=-V/2$.
At zero temperature, the states that contribute to the net current are the states of the
left lead with energies between $\mu_R=-V/2$ and $\mu_L=V/2$.
Evaluating (\ref{eq-30})-(\ref{eq-31}),
we arrive to the following expression for the current
\begin{equation}
I(V)=\frac{2e}{h}\int_{-V/2}^{V/2} T(\varepsilon)\;\mathrm{d}\varepsilon,
\end{equation}
where $T(\varepsilon)$ is the transmission. We have assumed that the Fermi level in equilibrium
is equal to $0$.

\section{Results}

The transmission probability can be written in the simple form
\begin{equation}
T(\varepsilon)=\frac{4t_c^2\Gamma^2(\Delta\varepsilon)^4\cos^2{\phi/2}}
{\left[(\Delta\varepsilon)^4+\Gamma^2(\varepsilon-q)^2\right]\left[(\Delta\varepsilon)^4
+\Gamma^2(\varepsilon+q)^2\right]} \label{eq-9}
\end{equation}
where $(\Delta\varepsilon)^2= \varepsilon^2-p^2$, with
$p^2=(V/2)^2+t_c^2$ and $q^2=(V/2)^2+t_c^2\cos^2{(\phi/2)}$, where
$\Gamma=4\pi t_A^2\rho(0)$ is
the characteristic line width, with $\rho(0)$ the density of
states in the leads at the Fermi level.
It is evident from the expression (\ref{eq-9}) that the transmission
has a period $\Delta\phi=2\pi$ ($\Delta \Phi=\Phi_0$).  On the other hand,
it is totally suppressed, both in and out of equilibrium,
when $\phi$ is an odd multiple of $\pi$ ($\Phi=n\Phi_0/2$, $n$ odd).
This fully destructive interference effect for this value of $\phi$ is expected,
since in the absence of magnetic flux the upper and lower paths are equivalent,
then the magnetic flux introduces in the wave function a phase $-\pi/2$ along one arm
and $\pi/2$ along the other.
The total suppression of transmission was discussed in an Aharonov-Bohm
interferometer with two quantum dots, in equilibrium \cite{kubala}.

\subsection{Equilibrium transport}

\begin{figure}[h]
  \centering
  \includegraphics[width=5cm,angle=-90]{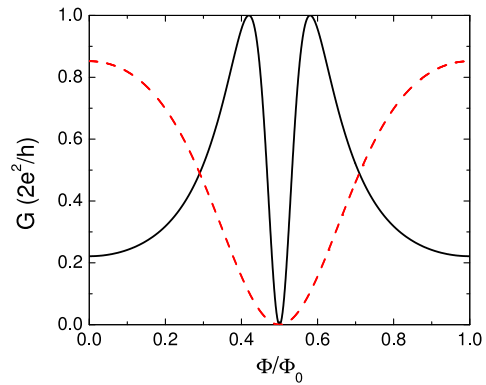}\\
  \caption{(Color online) Conductance versus magnetic flux for $t_c=0.25\Gamma$ (solid line)
  and $t_c=1.5\Gamma$ (dash line).}\label{condphi}
\end{figure}

The equilibrium transmission is obtained from
Eqs. (\ref{eq-9})-(\ref{eq-11}) by making $V=0$. Let us first analyze
the action of the magnetic flux in the conductance $G=(2e^2/h)T(0)$.
This is given by
\begin{equation}
G=\frac{2e^2}{h}\frac{4\cos^2(\pi\Phi/\Phi_0)(t_c/\Gamma)^2}
{\left[\cos^2{(\pi\Phi/\Phi_0)}+(t_c/\Gamma)^2\right]^2}.
\label{conduct}
\end{equation}
We distinguish in Eq. (\ref{conduct}) two different behaviors. When $t_c\leq \Gamma$ there is always a
value of $\Phi$ for which $G=2e^2/h$, while  when $t_c> \Gamma$ this never
occurs.  These behaviors are illustrated in Fig. \ref{condphi}, which shows
the conductance versus the
magnetic flux for two different values of the $t_c$.
For $t_c=0.25\Gamma$, the conductance reaches the maximum
$G=2e^2/h$ in the interval $\Phi=0$ to $\Phi=\Phi_0/2$, being
symmetric around $\Phi_0/2$. Such a maximum occurs at $\varepsilon=\arccos{(t_c/\Gamma)}\Phi_0/\pi$.
Similar features are found for any
value of $t_c\leq \Gamma$, no matter how small it is. This is a
remarkable result, taking into account that for a molecule in
series the conductance, proportional a $t_c^2$, gets progressively
smaller as $t_c$ decreases\cite{vanderwiel}. In contrast, for $t_c
> \Gamma$ the conductance decays monotonously with $\Phi$ in the
same interval, never reaching the quantum limit.

\begin{figure}[ht]
\centering
  \includegraphics[width=5.2cm,angle=-90]{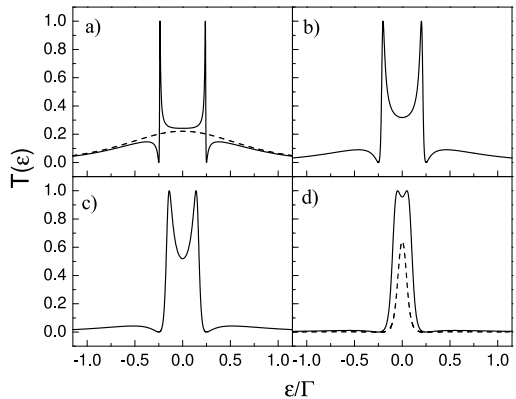}
  \caption{Transmission versus energy for $t_c=0.25\Gamma$ and a) $\Phi=0$ (dash line) and $\Phi=0.1\Phi_0$ (solid line), b) $\Phi=0.2\Phi_0$,
  c) $\Phi=0.3\Phi_0$,  d) $\Phi=0.4\Phi_0$ (solid line) and $\Phi=0.46\Phi_0$ (dash line). }
  \label{Fig3}
\end{figure}

The occurrence of maximum conductance for small values of $t_c$ is associated to an
effective level attraction of the hybridized levels,
similar to the discussed in Ref.
\onlinecite{kubala} in an Aharonov-Bohm interferometer with two
single-level quantum dots. This effect is illustrated in Fig.
\ref{Fig3}, which shows transmission versus energy for
$t_c=0.25\Gamma$.
%
In the absence of magnetic flux, the transmission behaves qualitatively as that of the DQD in series.
In Fig. \ref{Fig3}a (dash line), $t_c$ is small and a single and
flat peak is observed. The hybridized states are not resolved yet,
which is the usual for a molecule in series with small $t_c$.
However, two BICs are occurring, similarly to what
happen for a DQD embedded in an AB ring\cite{ghost,fanodicke}.
In other words, two of the hybridized states are localized and do not participate of
transmission. This situation repeats whenever $\Phi=n\Phi_0$ ($n$ integer). When
the flux is on, the BICs are replaced by two Fano resonances, as seen in the rest of Figs.
\ref{Fig3}. The Fano peaks reach $T=1$.

\begin{figure}[ht]
\centering
  \includegraphics[width=5.2cm,angle=-90]{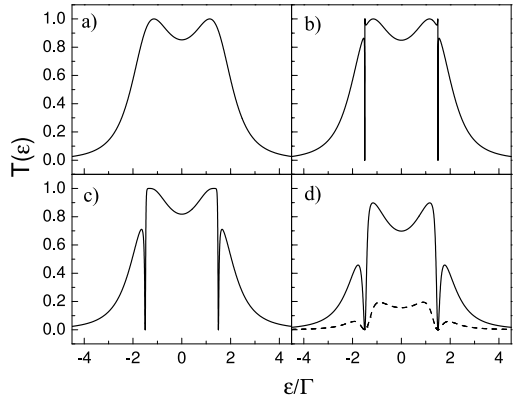}
  \caption{Transmission versus energy for $t_c=1.5\Gamma$ and a) $\Phi=0$, b) $\Phi=0.03\Phi_0$, c)  $\Phi=0.1\Phi_0$, d)  $\Phi=0.2\Phi_0$ (solid line) and $\Phi=0.4\Phi_0$ (dash line).}\label{Fig4}
\end{figure}

\begin{figure}[h]
  \includegraphics[width=5cm,angle=-90]{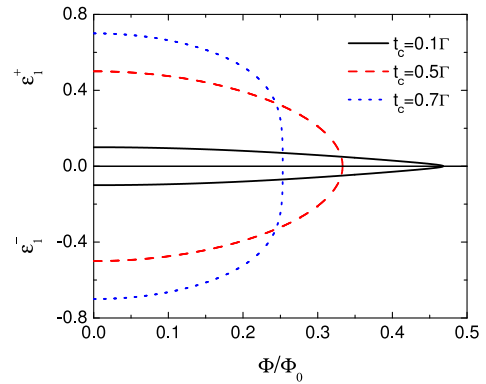}
  \caption{(Color online) Positions of the $T=1$ peaks for different values of $t_c\leq \Gamma/\sqrt{2}$.}\label{Fig-20}
\end{figure}

\begin{figure}[h]
   \includegraphics[width=5cm,angle=-90]{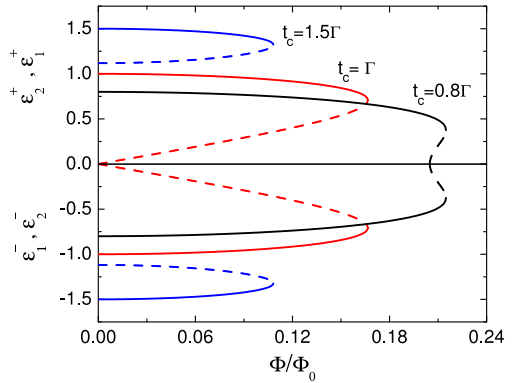}
  \caption{Positions of the maxima of transmission for different values of
  $t_c> \Gamma/\sqrt{2}$. The solid lines correspond to $\varepsilon_1^+$ and $\varepsilon_1^-$, and the dash lines to  $\varepsilon_2^+$ and $\varepsilon_2^-$.}\label{Fig-21}
\end{figure}

In the sequence from Fig. \ref{Fig3}a to Fig. \ref{Fig3}d it can
be observed the flux-controlled level attraction between hybridized
states. As $\Phi$ is increased from $0$ to $\Phi_0/2$ the Fano peaks get
progressively closer to each other until overlapping completely,
the transmission decaying when $\Phi$ is close to $\Phi_0/2$ (dash
line in Fig. \ref{Fig3}d), until vanishing.
The value $2e^2/h$ in
the conductance shown above is reached when the two peaks overlap
totally.
In contrast, for $t_c=1.5\Gamma$ the level
attraction is not significant, as observed in Fig. \ref{Fig4}.
\\
It follows from Eq. \ref{eq-9} that perfect transmission takes
place at energies obeying the following equation
\begin{equation}
\varepsilon^4-\varepsilon^2(2t_c^2-\Gamma^2)+t_c^2[t_c^2-\Gamma^2\cos^2(\phi/2)]=0,
\label{eq-11}
\end{equation}
with two pair of solutions
\begin{equation}
\varepsilon_{1}^\pm=\pm[A+B^{1/2}]^{1/2},
\quad\varepsilon_{2}^\pm=\pm[A-B^{1/2}]^{1/2}
\end{equation}
with $A=t_c^2-\Gamma^2/2$ and $B=\Gamma^2-4t_c^2\sin^2{(\phi/2)}$.
Fig. \ref{Fig-20} shows the positions of the $T=1$ peaks for different values
of $t_c\leq \Gamma/\sqrt{2}$. These correspond to $\varepsilon_1^+$ and $\varepsilon_1^-$, the only real solutions of Eq. (\ref{eq-11}). In the three cases the peaks positions shift progressively to the center
of the band as $\Phi$ increases. The  peaks  meet at $\varepsilon=0$ at
$\Phi=\arccos{(t_c/\Gamma)}\Phi_0/\pi$.
The situation slightly changes when $\Gamma/\sqrt{2}<t_c\leq\Gamma$, where two new real solutions
of Eq. (\ref{eq-11}) arise in the interval $\arccos{(t_c/\Gamma)}\Phi_0/\pi <\Phi \leq \arccos{(1-\Gamma^2/2t_c^2)}\Phi_0/2\pi$, while $\varepsilon_1^+$ and $\varepsilon_1^-$
become real for all $\Phi\leq \arccos{(1-\Gamma^2/2t_c^2)}\Phi_0/2\pi$. This is illustrated
in Fig. \ref{Fig-21} for $t_c=0.8\Gamma$ and $t_c=\Gamma$. The solutions $\varepsilon_1^+$ and
$\varepsilon_1^+$ never meet, but $\varepsilon_2^+$ and $\varepsilon_2^-$ do so at $\varepsilon=0$, being
responsible of the conductance $G=2e^2/h$ for these values of $t_c$.
When $t_c>\Gamma$ there are four $T=1$ peaks when $\Phi\leq \arccos{(1-\Gamma^2/2t_c^2)}\Phi_0/2\pi$. These peaks never reach the center of the band, as shown for $t_c=1.5\Gamma$.
\\
We show below that he flux-dependent level attraction and the complete overlapping of resonances
present for small values of $t_c$ strongly influence the behavior of the non-equilibrium transport.

\subsection{Non-equilibrium transport}

\begin{figure}[ht]
\centering
  \includegraphics[width=6cm,angle=0]{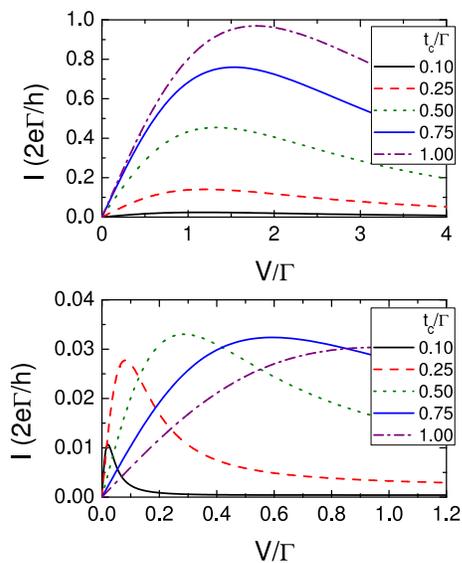}
  \caption{(Color online) $I$-$V$ characteristics for $\Phi=0$ (upper panel)
     and $\Phi=0.46\Phi_0$ (lower panel), and
   $t_c=0.1\Gamma$ (solid line), $t_c=0.25\Gamma$ (dash line), $t_c=0.5\Gamma$ (dotted line),
    $t_c=0.75\Gamma$ (dash-dotted line), $t_c=\Gamma$ (short dash line).} \label{Fig5}
\end{figure}

Let us now consider a voltage $V$ applied between contacts and let us study the current
in the leads.
Fig. \ref{Fig5} (upper panel) shows the current-voltage characteristics for
$\Phi=0$ and different values of intramolecular couplings.

In all cases the current-voltage characteristics displays a peak, with the corresponding region
of NDC.  This feature occurs analogously to the observed in the serial DQD\cite{aguado}.
The current increases when the bias allows a transmission channel exists along the
left and right sides of the system.
If the bias continues to increase, the channel is destroyed resulting in the drop of current.
On the other hand, as occurs for
quantum dots connected in series, in general larger values of $t_c$ give
larger values of the current for the same voltage.

\begin{figure}[ht]
\centering
 \includegraphics[width=5.5cm,angle=-90]{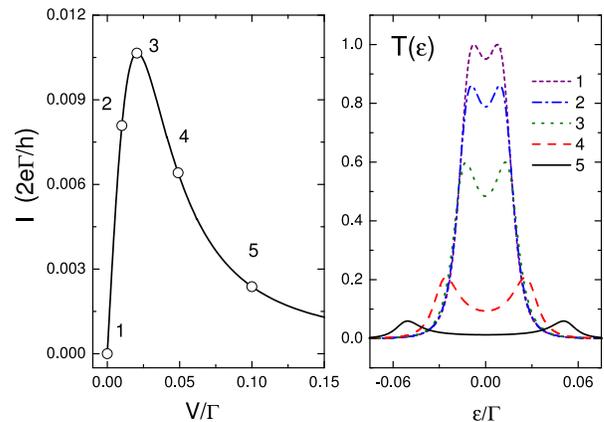}\\
  \caption{(Color online) $I$-$V$ characteristics for
   $t_c=0.1\Gamma$ and $\Phi=0.46\Phi_0$ (left panel). Transmission spectrum for different bias voltages
   for the same parameters (right panel). } \label{Fig6}
\end{figure}

\begin{figure}[ht]
\centering
 \includegraphics[width=5.5cm,angle=-90]{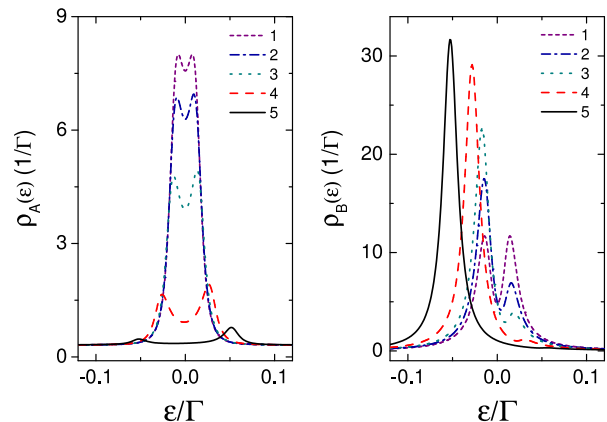}\\
  \caption{(Color online) Densities of states of the electrons coming from the left at the quantum dots A (left panel) and B (right panel), for
   $t_c=0.1\Gamma$ and $\Phi=0.46\Phi_0$.} \label{Fig9}
\end{figure}

The existence of a magnetic flux produces changes in the $I$-$V$
characteristics which become important when $t_c< \Gamma/\sqrt{2}$ and
$\Phi$ is within an interval close to $\Phi_0/2$,
as illustrated in
Fig. \ref{Fig5} (lower panel), where $\Phi=0.46\Phi_0$.
We observe sharper current peaks at lower voltages as compared to the case $\Phi=0$.
The abrupt increase of the current at low voltages for small $t_c$ is consequence of the
level attraction discussed for zero bias.
To visualize this we have plotted in Fig. \ref{Fig6}
the $I$-$V$ characteristics for a fixed
$t_c$ and $\Phi$ (left panel),
and the transmission spectra associated to the bias voltages indicated in the
current curve (right panel).
As observed, for small bias (cases 1-3) the transmission keeps large
in all the transport region ($-V/2< \varepsilon <V/2$), due to the existence of two
overlapped resonances close to each other.
Larger bias voltages make the heights of the resonances fall,
so that the transmission in all the window of transport becomes smaller, occurring the observed
decrease in the current.
A further insight of this is obtained through the density of states of the left and right quantum dots.
Fig. \ref{Fig9} shows the left and right quantum dots DOS for the same parameters of Fig. \ref{Fig6}. In equilibrium ($V=0$) a molecular state is formed. For the cases 2 and 3 the coherence is still preserved but for higher voltages (cases 4-5) the physical picture changes.  In these cases, the coherence between dots is lost, the electron is localized at the left quantum dots and the molecular bridge is broken.

\begin{figure}[ht]
\centering
 \includegraphics[width=5.8cm,angle=-90]{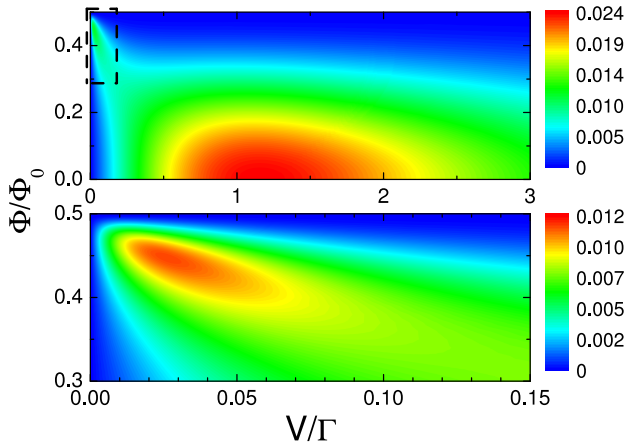}\\
  \caption{(Color online) Current versus bias voltage and magnetic flux for
   $t_c=0.1\Gamma$.} \label{Fig7}
\end{figure}

Fig. \ref{Fig7} shows the current versus bias voltage and magnetic flux for $t_c=0.1\Gamma$.
It can be identified clearly the two regions of maximal current at different scales.
In the upper panel, there is a broad peak centered in $\Phi=0$ and $V\approx1.2\Gamma$.
The behavior of the current versus voltage in this region was already
discussed for zero flux in Fig. \ref{Fig5}. The current maximum in this case decreases monotonously
with the magnetic flux, remaining its position almost unchanged.
In the same panel it is highlighted a region close to $\Phi_0/2$ and small voltages, which is
plotted in the lower panel. The observed peak corresponds to
the enhancement of current with the magnetic flux taking place for weak intramolecular
couplings, described in Fig. \ref{Fig5}, lower panel.
The maximum of current now occurs around $\Phi=0.44\Phi_0$ and $V=0.029\Gamma$.
Also, this peak is clearly defined when $t_c\leq \Gamma/\sqrt{2}$, when the flux produces a total
level attraction in equilibrium and it is not present when $t_c\sim \Gamma$ or larger.
It is important to note that the latter feature does not exist if the magnetic flux is absent, so that
in this case we can properly talk of magnetic flux-induced NDC.
Similar results are discussed in a parallel DQD molecule embedded in an Aharonov-Bohm ring\cite{dong} and
in a molecule in a three terminals configuration\cite{mourokh}, in both cases NDC occured in the strong interdot repulsion limit.

We expect that the above picture remains valid even if the
electron-electron interaction is taken into account. In fact, in
embedded QD arrays, the main effect of the electron-electron
interaction is to shift and to split the resonance positions\cite{chen,yu}.
This occurs because the on-site Coulomb repulsion energy U
introduces a renormalization of the site energies. In analogy
with  QD arrays in series, we expect that depending on the
relation between the interdot coupling and the on-site Coulomb
interaction different regimes arise. For $t_c /U \ll 1$, the
resonances and antiresonances would split into two distinct
minibands separated by the on-site Coulomb energy, while
for $t_c /U\gg 1$, the resonances and antiresonances would occur
in pairs.  We think that the above behavior would not break the negative differential 
conductance. A work in this direction is under progress.

\section{Summary}

We studied the transmission and the $I$-$V$ characteristics for two double quantum dot molecules embedded in  an Aharonov-Bohm ring. We showed that for $t_c\leq \Gamma$,
the magnetic flux can be used to control totally the conductance, allowing this to take any value between $0$ and $2e^2/h$.
When $t_c\leq \Gamma/\sqrt{2}$ the flux produces an effective level attraction and lines of perfect transmission, allowing the levels to meet at the center of the band at a determinate value of the flux.
On the other hand, the system displays BICs when $\Phi=n\Phi_0$ ($n$ integer),
and total suppression of transmission when $\Phi=n\Phi_0/2$ $n$ odd.
In the non-equilibrium regime, we identify two kind of NDC regions
in the $I$-$V$ characteristics, occurring at different scales and of different origin.
A first current peak exists at voltages of the order of the characteristic
linewidth $\Gamma$, and it is independent of the magnetic flux.
The role of the flux in this case is to control the height of
the overall current. The drop of current with the increase of bias has analogous explanation
to the NDC region in a serial DQD.
A second peak in the $I$-$V$ characteristics takes place only when $t_c<\Gamma$,
at voltages $V\sim\Gamma/10$ and it is strongly dependent on the magnetic flux. In fact,
it does not exist if the flux is absent. The current suffers an abrupt
rise for small bias voltages, as consequence of
an effective level attraction of the hybridized levels produced
by the flux. The decrease of current is result of the
destruction of this effect when the bias is increased.

\section*{Acknowledgments}

The authors acknowledge financial support from FONDECYT, under
grant 1080660. M. L. L. de G. thanks financial support from
Milenio ICM P06-067-F, and P. A. O. and G. A. L. from
CONICYT/Programa Bicentenario de Ciencia y Tecnolog\'{\i}a
(CENAVA, grant ACT27).


\begin{thebibliography}{99}




\bibitem{petta} J. R. Petta, A. C. Johnson, J. M. Taylor, E. A. Laird, A. Yacoby, M. D. Lukin, C. M. Marcus, M. P. Hanson, and A. C. Gossard, Science 309, 2180 (2005); A. K. H\"uttel, S. Ludwig, H. Lorenz, K. Eberl, and J. P. Kotthaus, Phys. Rev. B
\textbf{72}, 081310(R) (2005).

\bibitem{variousQD} L. I. Glazman, F. W. J. Hekking, and A. I. Larkin, Phys. Rev. Lett. \textbf{83}, 1830 (1999);
W.G. van der Wiel, S. De Franceschi, T. Fujisawa, J.M. Elzerman, S. Tarucha, and L.P. Kouwenhoven, Science \textbf{289}, 2105 (2000); T. H. Stievater, Xiaoqin Li, D. G. Steel, D. Gammon, D. S. Katzer, D. Park, C. Piermarocchi, and L. J. Sham, Phys. Rev. Lett. \textbf{87} 133603 (2001); T. Fujisawa, D. G. Austing, Y. Tokura, Y. Hirayama, and S. Tarucha, Nature \textbf{419}, 278 (2002).


\bibitem{ono} K. Ono, D. G. Austing, Y. Tokura, S. Tarucha, Science \textbf{297}, 1313 (2002).

\bibitem{vidan} A. Vidan, R. M. Westervelt, M. Stopa, M. Hanson, and C. Gossard, Appl. Phys. Lett. \textbf{85}, 3602 (2004).

\bibitem{amplif} P. Borri, S. Schneider, W. Langbein, U. Woggon, A. E. Zhukov, V. M. Ustinov, N. N. Ledentsov, Z. I. Alferov, D. Ouyang, and D.
Bimberg, Appl. Phys. Lett. \textbf{79}  2633 (2001);
A. V. Uskov, E. P. O'Reilly, M. Laemmlin, N. N. Ledentsov, and D.
Bimberg, Opt. Commun. \textbf{248}, 211 (2005).

\bibitem{laser} N. N. Ledentsov, V. M. Ustinov, A. Y. Egorov, A. E. Zhukov, M. V. Maksimov,
I. G.Tabatadze, and P. S. Kopev, Semiconductors \textbf{28}, 832 (1994).

\bibitem{vanderwiel} W. G. van der Wiel, S. De Franceschi, J. M. Elzerman, T. Fujisawa, S. Tarucha, and
L. P. Kouwenhoven, Rev. Mod. Phys. \textbf{75} 1 (2003).

\bibitem{vandervaart} N. C. van der Vaart, S. F. Godijn, Y. V. Nazarov, C. J. P. M. Harmans,
J. E. Mooij, L. W. Molenkamp, C. T. Foxon, Phys. Rev. Lett.
\textbf{74}, 4702 (1995); F. R. Waugh, M. J. Berry, C. H. Crouch,
C. Livermore, D. J. Mar, R. M. Westervelt, K. L. Campman, and A.
C. Gossard, Phys. Rev. B \textbf{53}, 1413 (1996).

\bibitem{holleitner} A. W. Holleitner, C. R. Decker, H. Qin, K. Eberl, and R. H. Blick,
Phys. Rev. Lett. \textbf{87}, 256802 (2001).

\bibitem{ghost} M. L. Ladr\'on de Guevara, F. Claro, and P. A.
Orellana, Phys. Rev. B \textbf{67} 195335 (2003).

\bibitem{kang} K. Kang and S. Y. Cho, J. Phys. Condens. Matter \textbf{16}, 117 (2004).

\bibitem{kim} T.-S. Kim and S. Hershfield, Phys. Rev. B \textbf{63}, 245326 (2001).

\bibitem{boese} D. Boese, W. Hofstetter, and H. Schoeller, Phys. Rev.
B \textbf{66}, 125315 (2002).




\bibitem{sigrist} M. Sigrist, T. Ihn, K. Ensslin, D. Loss, M. Reinwald, and W. Wegscheider,
Phys. Rev. Lett. \textbf{96}, 036804 (2006).

\bibitem{ihn} T. Ihn, M. Sigrist, K. Ensslin, W. Wegscheider, and M. Reinwald,
New Journal of Phys. \textbf{9}, 111 (2007).





\bibitem{fanodicke} P. A. Orellana, M. L. Ladr\'on de Guevara, and F. Claro, Phys. Rev. B \textbf{70} 233315 (2004).

\bibitem{moldoveanu} V. Moldoveanu, M. Tolea, A. Aldea, and B. Tanatar, Phys. Rev. B \textbf{71},
125338 (2005).

\bibitem{sztenkiel} D. Sztenkiel and R. Swirkowikcz, J. Phys.:
Condens. Matter \textbf{19}, 176202 (2007).

\bibitem{dong} B. Dong, X. L. Lei, and N. J. M. Horing, Phys. Rev. B
\textbf{77}, 085309 (2008).


\bibitem{ramsak}  A. Ramsak, J. Mravlje, R. Zitko, and J. Bonca, Phys. Rev. B \textbf{74},
241305(R) (2006).

\bibitem{sztenkiel2} D. Sztenkiel and R. Swirkowikcz, J. Phys.:
Condens. Matter \textbf{19}, 386224 (2007).

\bibitem{kubala} B. Kubala and J. K\"onig, Phys. Rev. B \textbf{65}, 245301 (2002).

\bibitem{apel} V. M. Apel, M. A. Davidovich, G. Chiappe, E. V. Anda, Phys. Rev. B \textbf{72}, 125302 (2005).

\bibitem{weis} J. Weis, R. J. Haug, K. v. Klitzing, and K. Ploog, Phys. Rev. Lett. \textbf{71}, 4019 (1993).

\bibitem{sollner} T. C. L. G. Sollner, P. E. Tannenwald, D. D. Peck and W. D. Goodhue, Appl. Phys. Lett. 45, 1319 (1984)

\bibitem{weinman} D. Weinmann, W. H\"ausler, and B. Kramer, Phys. Rev. Lett. \textbf{74}, 984 (1995).

\bibitem{ciorga} M. Ciorga, M. Pioro-Ladriere, P. Zawadzki, P. Hawrylak, and A. S. Sachrajda,
Appl. Phys. Lett. 80, 2177 (2002).

\bibitem{thielmann} A. Thielmann, M. H. Hettler, J. K\"onig, and G. Sch\"on, Phys. Rev. B
\textbf{71}, 045341 (2005).

\bibitem{rogge} M. C. Rogge, F. Cavaliere, M. Sassetti, R. J. Haug, and B. Kramer,
New J. Phys. \textbf{8} 298 (2006).



\bibitem{aguado} R. Aguado and D. C. Langreth, Phys. Rev. Lett.
\textbf{85}, 1946 (2000).

\bibitem{lara} G. A. Lara, P. A. Orellana, and E. V. Anda, Solid State Commun. \textbf{125}, 165 (2003).

\bibitem{fransson} J. Fransson and O. Eriksson, J. Phys.: Condens. Matter \textbf{16}, L85 (2004);
J. Fransson and O. Eriksson, Phys. Rev. \textbf{70}, 085301 (2004).

\bibitem{wunsch} B. Wunsch, M. Braun, J. K\"onig, and D. Pfannkuche, Phys. Rev. B \textbf{72},
205319 (2005).

\bibitem{nguyen} V. H. Nguyen, V. L. Nguyen, and P. Dollfus, Appl. Phys. Lett. \textbf{87}, 123107 (2005).

\bibitem{pedersen} J. N. Pedersen, B. Lassen, A. Wacker, and M. H. Hettler, Phys. Rev. B \textbf{75}, 235314 (2007).

\bibitem{mourokh} L. G. Mourokh and A. Y. Smirnov, Phys. Rev. B \textbf{72}, 033310 (2005).

\bibitem{lara2} G. A. Lara, P. A. Orellana, and E. V. Anda, Phys. Rev. B \textbf{78}, 045323 (2008).

\bibitem{li} Y.-X. Li, H.-Y. Choi, and H.-W. Lee, Phys. Lett. A \textbf{372}, 2073 (2008).


\end{thebibliography}
\end{document}